\documentclass[useAMS,usenatbib,fleqn]{mnras}
\usepackage{listings}
\usepackage{amsmath}
\usepackage{natbib}
\usepackage{graphicx}
\usepackage{color}
\usepackage{rotating}
\usepackage{nicefrac}
\usepackage{txfonts}
\usepackage{trivfloat}
\usepackage{subcaption}
\usepackage{adjustbox}
\trivfloat{listfloat}
\usepackage{longtable}
\usepackage[percent]{overpic}
\usepackage{xspace}
\usepackage{url}
\usepackage{mathrsfs}
\usepackage{amssymb}
\usepackage{mathtools} 
\usepackage{enumitem}
\usepackage[percent]{overpic}
\usepackage{soul}

\usepackage{epsfig}
\usepackage{amsmath}
\usepackage{rotating}
\usepackage{natbib}
\usepackage{multirow}
\usepackage[noend]{algpseudocode}

\lstdefinestyle{customc}{
  belowcaptionskip=1\baselineskip,
  breaklines=true,
  language=C,
  showstringspaces=false,
  basicstyle=\footnotesize\ttfamily,
}

\usepackage{array}
\usepackage{appendix}
\usepackage{comment}

\include{mnras_jdf}
\author[Nassif-Lachapelle \& Tamayo]{Loic Nassif-Lachapelle $^1$, 
    Daniel Tamayo$^{2}$\thanks{NHFP Sagan Fellow}\\
$^1$ Department of Astronomy and Astrophysics, University of Toronto, Toronto, Ontario, M5S 3H4, Canada\\
$^2$ {Department of Astrophysical Sciences, Princeton University, Princeton, New Jersey 08544, United States}\\
}
\title{Direct Imaging of Irregular Satellite Disks in Scattered Light}
\date{Draft version: \today{}}

\vspace{0.5\baselineskip}

\begin{document}
\maketitle

\begin{abstract}
Direct imaging surveys have found that long-period super-Jupiters are rare.
By contrast, recent modeling of the widespread gaps in protoplanetary disks revealed by ALMA suggests an abundant population of smaller Neptune to Jupiter-mass planets at large separations.
The thermal emission from such lower-mass planets is negligible at optical and near-infrared wavelengths, leaving only their weak signals in reflected light. 
Planets do not scatter enough light at these large orbital distances, but there is a natural way to enhance their reflecting area.
Each of the four giant planets in our solar system hosts swarms of dozens of irregular satellites, gravitationally captured planetesimals that fill their host planets' spheres of gravitational influence. 
What we see of them today are the leftovers of an intense collisional evolution.
At early times, they would have generated bright circumplanetary debris disks. We investigate the properties and detectability of such irregular satellite disks (ISDs) following models for their collisional evolution from \cite{Kennedy11}.
We find that the scattered light signals from such ISDs would peak in the $10-100$ AU semimajor axis range implied by ALMA, and can render planets detectable over a wide range of parameters with upcoming high-contrast instrumentation.
We argue that future instruments with wide fields of view could simultaneously characterize the atmospheres of known close-in planets, and reveal the population of long-period $\sim$ Neptune-Jupiter mass exoplanets inaccessible to other detection methods.
This provides a complementary and compelling science case that would elucidate the early lives of planetary systems.

\end{abstract}


\section{Introduction}

Submillimeter observations with the Atacama Large Millimeter Array (ALMA) have revealed that gaps and rings are common features in the brightest, nearby protoplanetary disks \citep{Andrews18}. 
Such structures can potentially be explained through a variety of (magneto)hydrodynamic effects \citep[e.g.,][]{Takahashi14, Flock15, Loren15, Bethune16} or condensation fronts \citep[e.g.,][]{Zhang15}. 
However, they are most often interpreted as signposts of the formation of giant planets \citep [e.g.,][]{Brogan15, DiPierro15, Tamayo15, Pinte15}.

By modeling the dust gaps opened by planets of various masses across a range of disk parameters, \cite{Zhang18} infer that most of the gaps observed by the DSHARP survey \citep{Andrews18} are consistent with planets with masses between that of Neptune and Jupiter ($M_N - M_J$). 
Strikingly, \cite{Zhang18} infer that the occurrence rate of such planets is $\sim 50\%$, spread log-uniformly from $\sim 10-200$ AU.
If correct, this presents at least two important problems for planet formation theory.

First, planet formation theory has difficulties forming giant planet cores at tens of AU separations from their host star within the $\sim$ Myr lifetimes of gas disks \citep[see][for a review]{Goldreich04}, though pebble accretion may help alleviate the tension \citep{Johansen17, Rosenthal18}. 
Second, even if giant planet cores can be formed, they should then undergo fast, runaway accretion to become gas giants once their gas envelope becomes comparable in mass to the solid core \citep[e.g.,][]{Pollack96, Piso14, Lee14}. 
Why then, would one find $\sim$ Neptune-mass planets in this short-lived runaway mass-range while gas remains available in the disk for accretion?

Given that a planetary origin for the gap structures found with ALMA would thus provide fundamental constraints on current planet formation theory, observational tests of this hypothesis are a critical effort in exoplanet science and are being vigorously pursued from multiple directions.

One way to do this during the disk phase is to look for perturbations to the disk's gas pressure profile from embedded planets. In this way, \cite{Teague18} and \cite{Pinte18} were able to indirectly detect the kinematic signature of Jupiter-mass planets embedded in the HD 163296 disk. Another is to detect planets through their much brighter {\it circumplanetary} disks in either scattered light \citep{Szulagyi19} or thermal emission \citep{Szulagyi18}. Such searches are currently underway. However, both methods are observationally limited to finding planets $\gtrsim 1 M_J$, which are more massive than the population suggested by ALMA.

A complementary approach is to search for such a population of distant giant planets following disk dispersal. This has long been the focus of ground-based direct imaging, yielding several of the most iconic exoplanet detections to date, like HR 8799 \citep{Marois08} and Beta Pictoris b \citep{Lagrange10}, and enabling the spectroscopic characterization of these planets' atmospheres \citep[e.g.,][]{Janson10, Bowler10, Chilcote14}.

However, the conclusion from extensive direct imaging surveys is that planets $\gtrsim 5 M_\text{Jup}$ are rare beyond a few tens of AU, occurring only around approximately $1\%$ of stars on average \citep{Bowler18}, though about ten times more frequently around A stars \citep{Nielsen19}. 

At face value this seems to contradict the $\sim 50\%$ planetary occurrence inferred by \cite{Zhang18}. However, their ALMA observations implicate a much lower-mass planet population than is currently accessible through direct imaging. 
Such a result seems plausible from microlensing \citep[e.g.][]{Gould10} and radial velocity \citep[e.g.,][]{Cumming08} detections in the 1-10 AU range, which suggest that lower-mass ice giants like Neptune are significantly more common than gas giants $\gtrsim 1 M_J$. 
However, probing this large, putative $M_N - M_J$ population at $>10 AU$ separations will require pushing direct imaging detections to significantly lower masses.

\subsection{Thermal Emission vs. Scattered Light}

At face value, the prospects for detecting the abundant population of {\it long-period} Neptunes implied by the gaps in the DSHARP survey \citep{Zhang18} seems bleak for ground-based direct imaging. 
Detections to date have relied on searching for young planets' thermal emission in the near-infrared.
This is in the Wien limit of the blackbody emission, where the emission falls off exponentially as the temperature decreases.
Thus, only the hottest, most massive tail of the planet distribution is detectable at early times, dropping sharply toward lower masses and higher ages (Sec.\:\ref{secscalings}). 

Next generation direct imaging instruments therefore plan to push toward lower-mass planets by moving to the mid-infrared, or by searching for planets in scattered light.
In the latter case, the fraction of starlight intercepted by a planet, and thus its scattered light signal, grows quadratically the closer the planet is to its host star.
The optimal giant planet targets at 0.1" separations from nearby stars (corresponding to $\lesssim 5 AU$) yield typical contrast ratios $\sim10^{-8}$ \citep{Traub14}. 
Larger ELTs will move toward even smaller angular separations with brighter planets, but none of these approaches would constrain the much fainter, population of {\it long-period, low-mass} Neptunes implied by the gaps in the DSHARP survey \citep{Zhang18}.

\begin{figure} 
 \centering \resizebox{0.99\columnwidth}{!}{\includegraphics{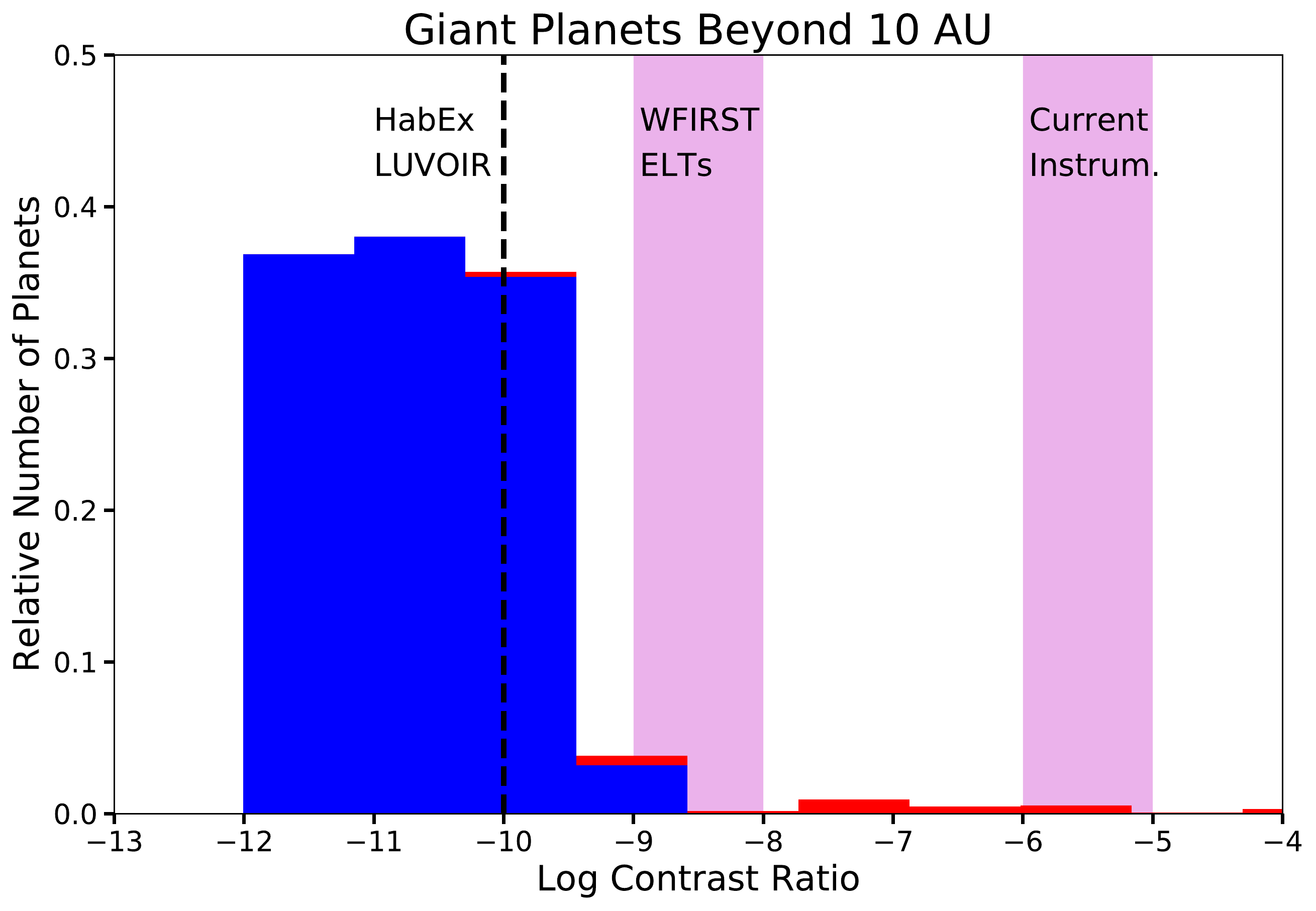}}
 \caption{
Contrast ratio histogram for the putative population of planets beyond 10 AU inferred from ALMA observations \citep{Zhang18}, with masses drawn from constraints by \protect\cite{Clanton16} combining direct imaging, microlensing and radial velocity detections.
Planets whose emission is dominated by thermal emission are colored red (all more massive than Jupiter), while planets that predominantly scatter incident starlight are in blue. 
Vertical bands denote fiducial contrast ratio detection thresholds for current and future instrumentation.
Most of these planets beyond 10 AU are undetectable for the foreseeable future. \label{planetcontrasts}} 
\end{figure}

In Fig.\:\ref{planetcontrasts}, we plot the contrast ratios for a population of planets with semimajor axes drawn log-uniformly from 10-200 AU \citep{Zhang18}, and masses drawn from a power-law $dN/d \log{M} \propto M^{-0.86}$ between $M_N$ and $13 M_J$, which \cite{Clanton16} inferred by combining direct imaging, microlensing and radial velocity detections. 
For the host stars we take an approximation (Sec.\:\ref{gpies}) to the stellar sample of GPIES, the Gemini Planet Imager Exoplanet Survey \citep{Nielsen19}. 
The planet fluxes are a combination of scattered light and thermal emission in H band ($\approx 1.65 \mu$m), using planetary models from \cite{Baraffe08} with a heavy element mass fraction $Z=0.02$. 
This provides optimistic estimates of scattered light from low-mass planets, since in reality they would have higher metal fractions and thus smaller radii and scattering area. 
Planets whose flux is dominated by this scattered light component are colored in blue, while those that dominantly emit thermally are colored red. 
All planets dominated by their thermal emission with contrasts $>10^{-8}$ have masses higher than Jupiter. 
Low metal fractions are more appropriate for such objects, but higher values of $Z$ would only increase contrast ratios by at most a factor of 2. 

On the right of Fig.\:\ref{planetcontrasts}, we highlight in magenta the rough detection threshold for current state-of-the-art instrumentation. The exact contrast depends on the particular instrument, stellar magnitude and the planet's angular separation from its host \citep[e.g.,][]{Ruffio17}. 
More extreme contrasts are achievable with the Hubble Space Telescope (HST) Space Telescope Imaging Spectrograph \citep{Kalas08, Debes19} at wide angular separations around bright nearby stars, but the plotted range is appropriate for the fainter M dwarfs we will later focus on.
We also mark contrast goals for next generation direct imaging instruments on WFIRST and 30-m class telescopes (ELTs), as well as the Habitable Exoplanet Explorer (HabEx) and LUVOIR mission concepts, which would launch in the late 2030s.

In summary, while future direct imaging instruments will find and characterize many close-in exoplanets, the prospects for directly imaging the large population of long-period, Neptune-Jupiter mass planets putatively revealed by ALMA seem bleak for the foreseeable future. 
The best way to improve their detectability is by increasing their surface area, and our own Solar System suggests a natural way to do that.

\subsection{Irregular Satellites} \label{irregulars}

In our own solar system, each of the four giant planets hosts large populations of small irregular satellites of a few to $\sim 100$ km in diameter, roughly filling their Hill spheres. The Hill radius of the Hill sphere is defined as,
\begin{equation}
    R_H = a_{pl}\bigg(\frac{M_{pl}}{3M_{*}}\bigg)^{1/3}
\end{equation}
$M_{pl}$ and $M_{*}$ are the planet and stellar mass, respectively \citep[e.g.,][]{Lissauer09}.

These are thought to be objects left over from the era of planet formation that were gravitationally captured early in the solar system's history.
This results in swarms of mutually inclined, crossing orbits.

These irregular satellites should have been sourced from the same heliocentric population that fed Jupiter's Trojans, caught $60^\circ$ ahead and behind the giant planet at its triangular Lagrange points.
The fact that the irregular satellites exhibit much shallower size distributions than the Trojans implies an intense collisional evolution, of which we only see the remains \citep{Bottke10}.
Indeed, by modeling this process, \cite{Bottke10} infer that the irregular satellites represent the most collisionally evolved population in the solar system.
This implies that at early times, the collisional debris from such irregular satellite disks (ISDs) \footnote{ISDs would fill the Hill sphere isotropically, except the Kozai mechanism removes objects on orbits with inclinations $\gtrsim 40^\circ$ relative to the planet's orbital plane \citep[see, e.g., Fig. 3 of][]{Kennedy11}, leaving behind a vertically extended disk.} could have been orders of magnitude brighter than the planet itself \citep{Kennedy11}.

Not only could this help fill in the census of Neptune-Jupiter mass planets at large separations, it would also shed light on the mechanism for irregular satellite capture.
Planetesimals encounter isolated planets on hyperbolic trajectories, and thus require a way to lose energy in order to become bound.
\cite{Pollack79} proposed that drag from a circumplanetary gas disk could provide the requisite dissipation.
\cite{Cuk04} then showed that this could plausibly explain the prograde group of irregular satellites at Jupiter.
However, gas drag continues to operate following capture, leading to eventual loss of these bodies.
In this picture, therefore, the surviving irregular satellites are the last generation that was captured near the end of the disk phase.

By contrast, passing planetesimals could also be captured during close encounters between giant planets, losing enough energy from one planet to remain bound to the other.
\cite{Nesvorny07} found that this process could efficiently capture irregular satellites in the dynamical instability scenario envisioned by \cite{Tsiganis05} for the early solar system.

While the dominant capture mechanism remains uncertain, both the above processes are generic in planetary systems. 
Large planets must accrete in gaseous environments, and the distribution of giant exoplanets' orbital eccentricities is well reproduced by planetary close encounters \citep{Chatterjee08, Juric08}.

Detecting a sample of ISDs would not only help differentiate between these scenarios, it would also elucidate the early stages of planetary system assembly.
Most directly, discovery of gaseous giant planets at long orbital periods challenges theories to form them before disk dispersal.
This may point to processes such as pebble accretion that can speed up growth \citep{Lambrechts12, Rosenthal18}.
Additionally, the rain of ISD debris onto the central accreting protoplanet atmosphere could provide a significant opacity source, which would act to delay cooling and stall accretion \citep{Piso14, Lee15, BrouwersOrmel9}.
This could help explain how large numbers of Neptunes managed to accrete some gas without running away to form Jovian planets. 
Alternatively, if irregular satellites are captured during planetary encounters, ISDs would provide signposts for the timing of instabilities in young planetary systems.

This motivates investigating the collisional evolution of ISDs, both in order to inform their expected brightnesses, and the optimal observational strategies to find them.
\cite{Kennedy11} (henceforth K11) developed such models, applying them both to our own solar system's ISDs and the directly imaged object Fomalhaut b \citep{Kalas08, Kalas13}.
They also considered prospects for space-based follow-up with the Hubble and James Webb Space Telescopes.
Here we instead focus on their implications for observations with ground-based instruments, as well as WFIRST and the HabEx and LUVOIR mission concepts.

In Sec.\:\ref{model} we recapitulate the model of K11.
We then combine these results into analytic estimates of expected ISD contrast ratios in Sec.\:\ref{secscalings}, discuss their implications for observational strategies, and describe their observational signatures.
In Sec.\:\ref{projections} we make projections for current and future instrumentation using a target sample optimized for such ISD detections.
We summarize and conclude in Sec.\:\ref{conclusion}.

\section{Model} \label{model}

\subsection{Contrast Ratios}
The contrast ratio at wavelength $\lambda$ of starlight scattered off an astrophysical object is simply the fraction of light intercepted by the object's geometrical cross-section $\sigma$, multiplied by its single-scattering albedo $Q$ and phase function $g$ \citep[e.g.,][]{Collier02},
\begin{equation} \label{cr}
C(\lambda) = g(\alpha, \lambda)Q(\lambda)\frac{\sigma}{\pi a^2},
\end{equation}
where $a$ is the planet's semimajor axis (for simplicity we assume circular orbits), and $g$ is additionally a function of the phase angle of observation $\alpha$. 
Following \cite{Collier02} and K11, we set $g$ to a nominal value of $0.32$, the value for a Lambert sphere at maximum elongation from its host star.
Specializing to ISDs, we take the single-scattering Bond albedo of $\approx0.1$ measured for dust grains in Saturn's ISD, the Phoebe ring \citep{Tamayo14, Tamayo16Phoebe},
\begin{equation} \label{contrast}
C \approx 1.0 \times 10^{-12} \Bigg(\frac{g}{0.32}\Bigg)\Bigg(\frac{Q}{0.1}\Bigg)\Bigg(\frac{90\text{AU}}{a}\Bigg)^2\Bigg(\frac{\sigma}{\pi R_J^2}\Bigg).
\end{equation}
The normalization of $\sigma$ to Jupiter's cross-sectional ($\pi R_J^2$) area shows that detecting a Jupiter directly in scattered light at tens of AU from its host star is far beyond the reach of current or planned technology.
However, an ISD can boost the scattering surface area $\sigma$ by orders of magnitude.

\subsection{Area-To-Mass Ratio}  \label{secAMratio}
We assume a steady state collisional cascade with the number of satellites $n(D)dD$ with diameter between $D$ and $D+dD$ given by $n(D) \propto D^{2-3q}$. Following K11, we consider a broken power law separating small bodies, which are dominantly held together by their material strength, from large bodies held together by self-gravity. In particular, we take a  power-law index $q=q_s$ from size $D_{min}$ to $D_{t}$, and $q=q_g$ from size $D_t$ to the maximum satellite size that participates in the cascade $D_{c}$.

For typical particle size distributions, most of the surface area is in the smallest grains, while most of the mass is in the largest bodies. We will assume this throughout the paper. The disk brightness, set by the rate at which the collisional cascade is fed, is therefore determined by the mass liberated through collisions of the largest moons of size $D_c$ \citep{Wyatt07}. It is therefore instructive to express the surface area in terms of the disk mass $M$ through the cascade's area-to-mass ratio $\xi$, so $\sigma(t) = \xi M(t)$.

Because the initial mass and surface area are both simply proportional to the normalization of $n(D)$ above, they depend linearly on one another. Thus, the disk's area-to-mass ratio $\xi$ is independent of both time (assuming $D_c$ remains constant, see K11) and the total amount of material, and is set only by the parameters of the cascade.
To give concrete scalings, we take K11's nominal parameters of $q_s=1.9$, $q_g = 1.7$, in which case,
\begin{equation} \label{AMratio}
\Bigg(\frac{\xi}{\pi R_J^2/M_\oplus}\Bigg) \approx 5.5 \times 10^6 \Bigg(\frac{\rho}{\text{g/cc}}\Bigg)^{-1}\Bigg(\frac{D_\text{c}}{100\text{ km}}\Bigg)^{-0.9}\Bigg(\frac{D_t}{100\text{ m}}\Bigg)^{0.6}\Bigg(\frac{D_{min}}{\text{$\mu$m}}\Bigg)^{-0.7},
\end{equation}
where $M_{\oplus}$ is the mass of the Earth.
For a fixed initial mass budget $M_0$ at time $t=0$, one can thus increase $\sigma$ (through $\xi$) by either decreasing $D_\text{c}$, liberating mass in the small area-filling particles, or by decreasing $D_\text{min}$ and similarly increasing the number of small grains.

For the nominal parameters above, this implies that a $1 M_\oplus$ collisional cascade would boost a planet's contrast ratio by almost 7 orders of magnitude; we comment further on reasonable mass scales in Sec.\:\ref{projections}.
Finally, while the nominal parameter values in Eq.\:\ref{AMratio} are reasonable for all the solar system's irregular satellite populations, the minimum particle size $D_{min}$ will vary substantially with stellar mass.

The minimum particle size is set by radiation pressure moving small grains onto near-radial circumplanetary orbits, and depends on their orbital distance around the planet \citep{Burns79}.
Following K11, we take a characteristic separation from the planet $\eta$, expressed as a fraction of the planet's Hill sphere.
In that case, the minimum grain diameter is independent of the planet's semimajor axis, and given by (K11)
\begin{equation} \label{Dmin}
\frac{D_{min}}{\mu\text{m}} \approx 2\: \Bigg(\frac{\eta}{0.3}\Bigg)^{1/2} \Bigg(\frac{M_\star}{0.5 M_\odot}\Bigg)^{10/3}\Bigg(\frac{M_{p}}{M_J}\Bigg)^{-1/3}\Bigg(\frac{\rho}{\text{g/cc}}\Bigg)^{-1},
\end{equation}
where we have used that $L_\star/L_\odot \sim (M_\star/M_\odot)^4$ \citep{Popper80} to emphasize the strong dependence of radiation pressure on stellar mass. 

Equations \ref{AMratio} and \ref{Dmin} suggest that one can continue increasing the surface area of circumplanetary swarms by moving toward lower-mass stars and thus decreasing $D_{min}$. 
However, one eventually reaches a floor where $D_{min}$ becomes comparable to the wavelength of observation (e.g., $\approx 1.65 \mu$m when observing in H band with GPI).
At this point, observing even lower-mass stars no longer helps, since even though smaller, higher surface-area grains might exist, they can no longer effectively scatter photons at the wavelength of observation.
When observing in the near-infared, using the nominal parameters in Eq.\:\ref{Dmin}, this transition occurs at $M_\star \approx 0.5 M_\odot$, or roughly M1 stars.
We therefore normalize Eq.\:\ref{Dmin} at this optimum target mass, noting that our subsequent equations are only valid for $M_\star > 0.5 M_\odot$.
This is also roughly applicable even at shorter observation wavelengths, since although the observation wavelength would suggest targeting lower-mass stars, one would instead become limited by the star's spectrum peaking at longer wavelengths.

\subsection{Collisional Timescales} \label{secTcol}
Following \cite{Wyatt10}, K11 estimate a collision rate for the largest bodies of size $D_c$ that participate in and feed the cascade (Sec.\:\ref{timeevol}) through a particle-in-a-box formalism.
In this approximation, the rate of collisions for a moon of diameter $D_c$ is the product of the number density of impactors and the rate at which it sweeps out volume $\pi D_c^2 v_\text{rel}$, where $v_\text{rel}$ is a characteristic relative velocity. Throughout this paper we set this relative speed to the circular Keplerian velocity at a semimajor axis $\eta R_H$ around the planet, times a factor of $4/\pi$ as appropriate for a swarm of circular, isotropic orbits. Using more sophisticated methods, K11 estimate that this analytical result is correct within $\approx 30\%$ for typical parameters. Finally, K11 consider that the energy per target mass needed to shatter and disperse a body of size $D_c$ with gravity-dominated strength, is
\begin{equation} \label{QD}
\frac{Q_D^*}{\text{J/kg}} = \frac{3.3 \times 10^4}{f_Q} \bigg(\frac{\rho}{\text{g/cc}}\bigg)\bigg(\frac{D_c}{100 \text{ km}}\bigg)^{1.26},
\end{equation}
where $f_Q$ is a factor parametrizing the uncertainty in this quantity, which can vary by about an order of magnitude for various collision geometries and between different studies \citep[e.g.][]{Benz99, Leinhardt09}. \cite{Bottke10} find $f_Q > 3$ best fit the solar system irregular satellites, and K11 adopted $f_Q = 5$. Considering only collisions with bodies in the cascade with sufficient kinetic energy to disrupt a moon of size $D_c$ yields a collisional timescale for the largest bodies $T_{col}$ inversely proportional to the number of potential impactors and thus to the total mass $M$ of the ISD (K11),
\begin{equation} \label{Tcol}
\frac{T_{col}}{10\:\text{Myr}} = f_a^{-4.13}\:\bigg(\frac{M_p}{M_J}\bigg)^{0.24} \bigg(\frac{M_\star}{0.5\:M_{\odot}}\bigg)^{-1.38} \Bigg(\frac{a}{90\:\text{AU}}\Bigg)^{4.13}\Bigg(\frac{M}{M_\oplus}\Bigg)^{-1}, 
\end{equation}
where the swarm-specific parameters have been collected in the dimensionless parameter $f_a$ defined as
\begin{equation}
f_a \approx \Bigg(\frac{f_Q}{5}\Bigg)^{0.15} \Bigg(\frac{\eta}{0.3}\Bigg)^{-1} \bigg(\frac{\rho}{\text{g/cc}}\bigg)^{-0.39} \bigg(\frac{D_c}{100 \text{ km}}\bigg)^{-0.43}. \label{f_a}
\end{equation}
The reference values have been chosen from the solar system irregular satellites, so $f_a$ should be of order unity, and its exponent in Eq.\:\ref{Tcol} was chosen to simplify the interpretation of later expressions.

We briefly note several trends (for a deeper discussion see K11).
The strongest dependence is on $a$, largely since the Hill sphere's volume that must be explored for collisions to occur is proportional to $a^3$.
More subtly, if bodies orbit within a fixed fraction $\eta$ of the Hill sphere, the relative velocities get slower with increasing semimajor axis of the planet, reducing the ability to break up large bodies.
Collision times also get longer the smaller the available mass of impactors in the ISD $M$, so collision timescales will grow as the ISD grinds down with time.
There is also a near-inverse dependence on the stellar mass, largely because the Hill sphere's volume that must be explored for a collision is inversely proportional to $M_\star$.

For the swarm parameters encoded in $f_a$, the strongest dependence is on the size of the largest bodies in the collisional cascade $D_c$. 
The larger $D_c$ is, the fewer such bodies there can be for a fixed swarm mass, so it's harder for potential destructors to find these bodies as they wander the planet's Hill sphere.
It is also harder to break up larger bodies because self-gravity better holds them together.

Finally, as K11 note, there is a very weak dependence on planet mass, with important implications that we discuss in Sec.\:\ref{thermalcomp}.
While increasing the planet mass linearly increases the Hill sphere's volume in which collisions occur, it also speeds up the rate at which this phase space is explored by shortening the orbital periods.
Additionally, the larger relative velocities make it possible for comparatively smaller impactors to disrupt the largest bodies in the cascade.

\subsection{Mass time evolution} \label{timeevol}

Following K11, we assume that the size distribution remains fixed and that mass is lost from the cascade by the breakup of the largest objects of size $D_c$, in which case \citep{Wyatt07},
\begin{equation} \label{Mdot}
\dot{M} = -\frac{M}{T_\text{col}} \propto -M^2,
\end{equation}
since $T_\text{col} \propto M^{-1}$. One can verify through direct substitution that the solution is given by \citep{Wyatt07},
\begin{equation} \label{M(t)}
M(t) = \frac{M_0}{1 + t/T_\text{col}(M_0)},
\end{equation}
where $M_0$ is the initial mass of the ISD.
Thus, for $t \ll T_\text{col}(M_0)$, the mass remains fixed at the initial value because not enough time has elapsed for collisions. We will refer to such a regime as an age-limited disk, since the breakup of the largest bodies that feed the cascade is limited by the age of the system.

For $t \gg T_\text{col}(M_0)$, one can ignore the sum in the denominator, and the factor of $M_0^{-1}$ in $T_\text{col}$ (Eq.\:\ref{Tcol}) cancels with the numerator to yield $M_\text{tot}(t) \propto t^{-1}$, {\it independent of the initial mass} \citep{Wyatt07}. Physically, this is because the mass-dependence of the collision timescale (Eq.\:\ref{Tcol}) regulates the rate of grind-down. More massive disks collide and grind down faster, while less massive disks `wait' for their more massive counterparts to catch up, leading different initial masses to converge to the same late-time evolution (Fig.\:\ref{mvssep}). Following \cite{Heng10}, we will refer to this regime as a collision-limited disk.

Given that the strongest dependence is on the semimajor axis (Eq.\:\ref{Tcol}), K11 calculate the optimum value of $a$ at which to find an ISD, fixing all other parameters. 
This corresponds to finding the transition between the age-limited and collision-limited regimes, where the largest bodies have only just started colliding and haven't substantially ground down, i.e. where the collision time $T_{col}$ is equal to the age of the system $T_{age}$.
Inverting Eq.\:\ref{Tcol} then yields an optimum semimajor axis $a_{opt}$,
\begin{equation} \label{aopt}
\frac{a_{opt}}{90\: AU} = f_a \:\bigg(\frac{M_p}{M_J}\bigg)^{-0.06}\bigg(\frac{M_\star}{0.5\:M_{\odot}}\bigg)^{0.33}\Bigg(\frac{M_0}{M_\oplus}\Bigg)^{0.24}\Bigg(\frac{T_{age}}{10\:\rm{Myr}}\Bigg)^{0.24}.
\end{equation}
Thus, $f_a$ can be interpreted as the order-unity factor by which the swarm-specific parameters (Eq.\:\ref{f_a}) change the optimimum semimajor axis at which the ISD contrast ratio peaks.

An intuitive approximation of the mass evolution is that for $a>a_{opt}$, the largest bodies have not yet had time to disrupt, so $M \approx M_0$, while for $a<a_{opt}$ the debris disk has converged to a state in which the collision time of the largest bodies is equal to the system's age\footnote{This can be seen quantitatively by setting $T_\text{col} = t$ in Eq.\:\ref{Mdot}, which yields the same late-time solution $M \propto t^{-1}$.} \citep{Heng10}, and the remaining mass is independent of $M_0$ (Fig.\:\ref{mvssep}). 
Combining Eqs.\:\ref{Tcol} and \ref{M(t)},
\begin{equation} \label{Msplit}
\Bigg(\frac{M}{M_\oplus}\Bigg) \approx \left\{
        \begin{array}{ll}
            f_a^{-4.13}\:\bigg(\frac{M_p}{M_J}\bigg)^{0.24}\bigg(\frac{M_\star}{0.5\:M_{\odot}}\bigg)^{-1.38} \Bigg(\frac{a}{90\:\text{AU}}\Bigg)^{4.13}\Bigg(\frac{10\:\rm{Myr}}{T_{age}}\Bigg)  & \quad a < a_{opt} \\
            M_0/M_\oplus & \quad a > a_{opt}
        \end{array}
    \right.
\end{equation}

As a nominal example, we take an ISD around a Jupiter-mass planet around TWA 13A, an $\approx 8$th magnitude M1 star in H band, at $\approx 55$ pc in the TWA association \citep{Gagne17TWA}, which is estimated to be $\approx 7.5$ Myr old \citep{Ducourant14, Donaldson16}. 
We assume a fiducial stellar mass of $0.5 M_\odot$, a luminosity of $0.06 L_\odot$, and the nominal ISD parameters given in Eq.\:\ref{f_a}.
We plot in Fig.\:\ref{mvssep} the remaining ISD mass around planets of different semimajor axes for 4 different initial ISD masses (blue curves) of $1, 10^{-2}, 10^{-4}$ and $10^{-6} M_\oplus$.

The dashed red line is the approximation for $a<a_{opt}$ in the collision-limited regime (Eq.\:\ref{Msplit}), to which all initial masses converge. 
The mass estimate in this collision-limited regime is therefore particularly simple.
For example, at a semimajor axis of 10 AU, all initial masses converge to the same value of $\approx 10^{-4}M_\oplus$ in Fig.\:\ref{mvssep}, unless the initial mass was smaller than this value.
This would mean that there hasn't been enough time for the largest bodies in this swarm to collide, and the ISD mass remains at its initial value (e.g., the bottom curve, which remains at its initial value of $10^{-6} M_\oplus$).

\section{Contrast Ratio Scalings and Observational Signatures} \label{secscalings}

\begin{figure}
 \centering \resizebox{0.99\columnwidth}{!}{\includegraphics{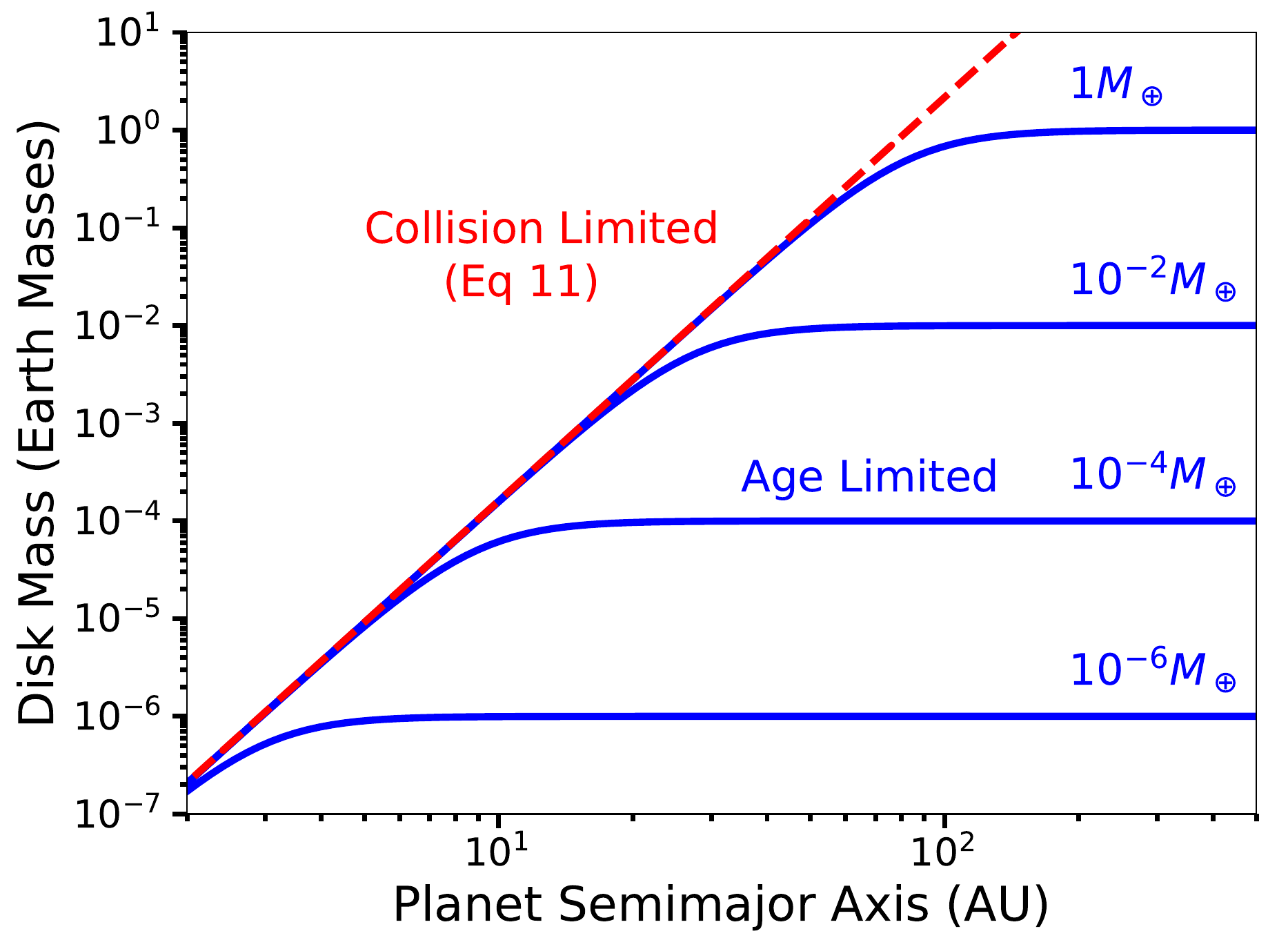}}
 \caption{
Remaining disk mass as a function of semimajor axis for four different initial ISD masses, at a fixed system age (blue curves, with stellar and ISD parameters given in text).
At wide separations in the age-limited regime, the largest bodies in the ISD have not yet had time to collide, and the masses remain at their initial values.
At closer separations, i.e., after many collision timescales, all curves converge onto the collision-limited approximation (dashed red line), so the mass becomes independent of the initial mass and given by the second case in Eq.\:\ref{Msplit}.
The transition from the age-limited to collision-limited regime happens at different locations $a_{opt}$ for different initial masses, given through Eq.\:\ref{aopt}.
\label{mvssep}}
\end{figure}

The strongest dependence of the contrast ratio is on the planet's semimajor axis, since that strongly influences the number of collision timescales the ISD has experienced and how much it has ground down.
So despite the fact that the fraction of intercepted starlight falls off quadratically with distance (Eq.\:\ref{cr}), the remaining mass in the collision-limited regime increases with semimajor axis even more steeply (Eq.\:\ref{Msplit}).
Thus, the contrast ratio will grow with semimajor axis until it reaches $a_{opt}$, beyond which the mass remains at its initial value (Fig.\:\ref{mvssep}), and the contrast then simply falls off quadratically with distance,

\begin{equation} \label{CRsplit}
C \approx \left\{
        \begin{array}{ll}
            C_{max} \Bigg(\frac{a}{a_{opt}}\Bigg)^{2.13} & \quad a < a_{opt} \\
            C_{max} \Bigg(\frac{a}{a_{opt}}\Bigg)^{-2} & \quad a > a_{opt}
        \end{array}
    \right.
\end{equation}

We can calculate the maximum contrast ratio $C_{max}$ by combining Eqs\:\ref{cr} and \ref{AMratio}, assuming $M=M_0$ at  $a=a_{opt}$. This yields

\begin{equation} \label{Cmax}
C_{max} = 3.3 \times 10^{-6}\:f_{Cmax}\:\Bigg(\frac{M_p}{M_J}\Bigg)^{0.34} \Bigg(\frac{M_\star}{0.5\:M_\odot}\Bigg)^{-3.0}\Bigg( \frac{T_{age}}{10 \text{Myr}}\Bigg)^{-0.48}\Bigg(\frac{M_0}{M_\oplus}\Bigg)^{0.52},
\end{equation}
with swarm-specific parameters collected in $f_{Cmax}$,
\begin{equation} \label{f_Cmax}
f_{Cmax} = \Bigg(\frac{g}{0.32}\Bigg)\Bigg(\frac{Q}{0.1}\Bigg)\Bigg(\frac{f_Q}{5}\Bigg)^{-0.31}\Bigg(\frac{\rho}{\text{g/cc}} \Bigg)^{0.5}\Bigg(\frac{D_t}{100 \text{m}}\Bigg)^{0.6}\Bigg(\frac{D_c}{100 \text{km}}\Bigg)^{-0.03}\Bigg(\frac{\eta}{0.3}\Bigg)^{1.65}.
\end{equation}

\begin{figure}
 \centering \resizebox{0.99\columnwidth}{!}{\includegraphics{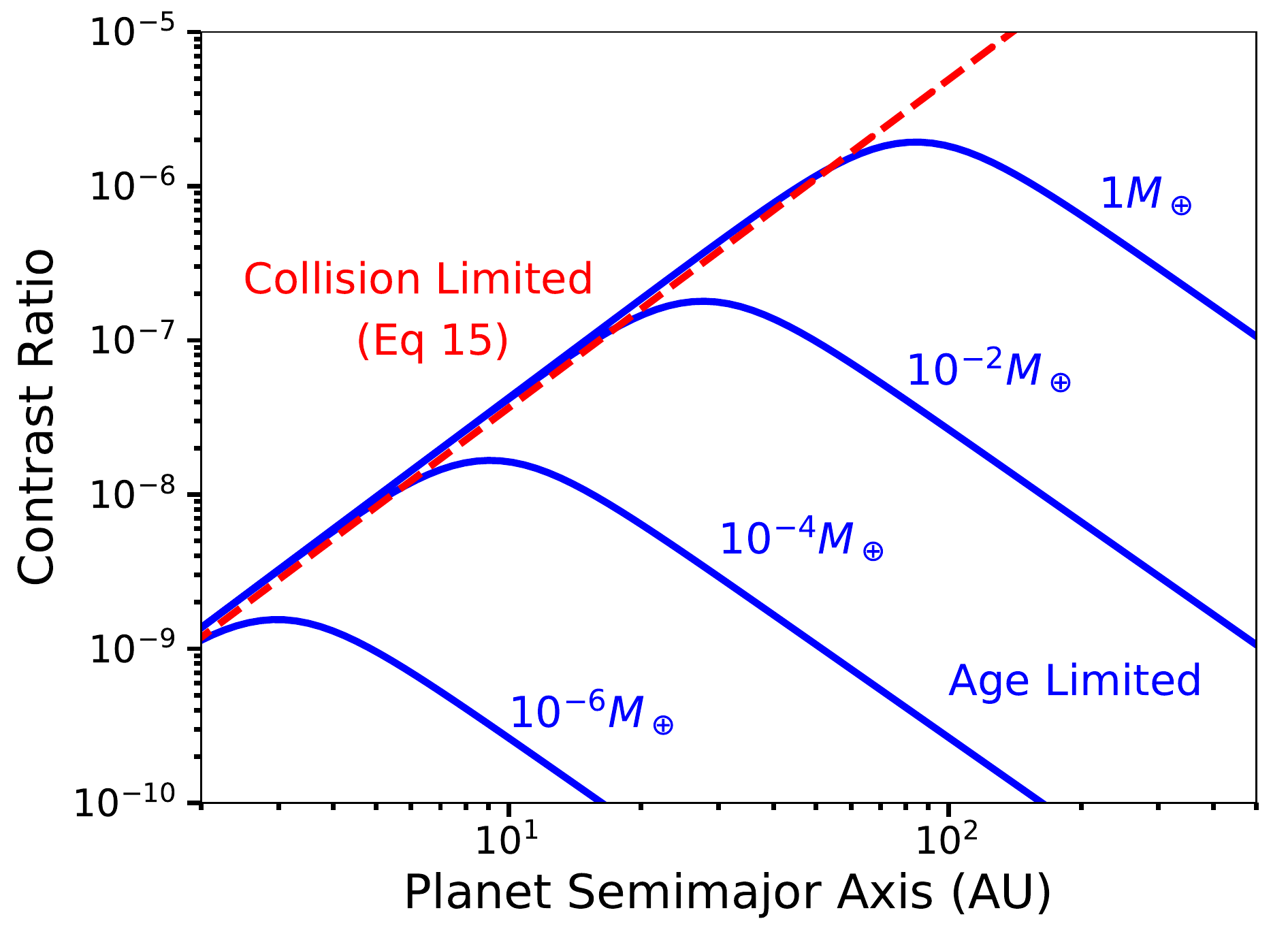}}
 \caption{
Contrast ratios as a function of semimajor axis for the same ISDs and system age as in Fig.\:\ref{mvssep}.
At large separations, in the age-limited regime, the mass remains constant (Fig.\:\ref{mvssep}), so contrasts fall off quadratically as the fraction of intercepted starlight.
At smaller separations, all tracks converge onto the same late-time evolution given by Eq.\:\ref{CRfinal} (red dashed line).
\label{CRvssep}}
\end{figure}

We plot exact solutions to the collisional model of K11 in blue lines in Fig.\:\ref{CRvssep} for the same initial masses as Fig.\:\ref{mvssep}, together with the top branch of Eq.\:\ref{CRsplit} as a dashed red line.
The exact solutions peak at $C_{max}/2$, since at $a=a_{opt}$, $t=T_{col}$ by definition, so $M_0$ has dropped by a factor of two through Eq.\:\ref{M(t)}.

We note that $C_{max}$ depends on the initial mass in irregular satellites $M_0$ (the peaks in Fig.\:\ref{CRvssep}  scale as $\approx M_0^{0.5}$ through Eq.\:\ref{Cmax}), because that sets how far out the collision-limited regime extends in Fig.\:\ref{CRvssep} before reaching $a_{opt}$ through Eq.\:\ref{aopt}.
But in the collision-limited regime, this dependence cancels out  as it should.
Plugging Eq.\:\ref{Cmax} and \ref{aopt} into Eq.\:\ref{CRsplit} yields for the collision-limited regime ($a<a_{opt}$) 

\begin{equation} \label{CRfinal}
C = 3.3 \times 10^{-6}f_C\:\Bigg(\frac{M_p}{M_J}\Bigg)^{0.47} \Bigg(\frac{M_\star}{0.5\:M_\odot}\Bigg)^{-3.71}\Bigg(\frac{a}{90 \text{AU}}\Bigg)^{2.13}\Bigg(\frac{10 \text{ Myr}}{T_{age}}\Bigg),
\end{equation}
with $f_C$,
\begin{equation} \label{f_C}
f_C =  \Bigg(\frac{g}{0.32}\Bigg)\Bigg(\frac{Q}{0.1}\Bigg)\Bigg(\frac{f_Q}{5}\Bigg)^{-0.63}\Bigg(\frac{\rho}{\text{g/cc}} \Bigg)^{1.33}\Bigg(\frac{D_t}{100 \text{m}}\Bigg)^{0.6}\Bigg(\frac{D_c}{100 \text{km}}\Bigg)^{0.89}\Bigg(\frac{\eta}{0.3}\Bigg)^{3.78}.
\end{equation}

Putting everything together, it is interesting to note that the parameter $f_{Cmax}$ (Eq.\:\ref{f_Cmax}) is fairly insensitive to its constituent swarm parameters.
This is because the two other combinations $f_a$ (Eq.\:\ref{f_a}) and $f_C$ (Eq.\:\ref{f_C}) are anti-correlated.
Swarm parameters that raise $f_C$ and move all the curves in Fig.\:\ref{CRvssep} upward tend to reduce $f_a$, so the blue curves peak at smaller semimajor axes, which has the opposite effect of lowering contrasts.
This renders contrast projections reasonably consistent despite  the large number of uncertain parameters. 
We quantify these expected variations over reasonable ranges of parameters in Sec.\:\ref{projections}.

In particular, the dependence on $D_c$ largely falls out in $f_{Cmax}$ (Eq.\:\ref{f_Cmax}), which sets the maximum contrast at $a=a_{opt}$.
The reason is that increasing $D_c$ (for a fixed total mass) acts to lock more of the mass in the largest bodies, decreasing the cross-sectional area of both the largest bodies (which sets $T_{col}$) {\it and} the smallest grains that dominate the total scattering surface area.
While the decrease in the latter will reduce the contrast ratio, the smaller cross-section for collisions in the largest bodies extends the collision times, and moves $a_{opt}$ inward (Eqs.\:\ref{aopt} and \ref{f_a}).
Because $C_{max}$ is by definition the contrast ratio at $a=a_{opt}$, it grows moving closer in to the star, where the total mass reflects a higher proportion of the stellar flux (Eq.\:\ref{contrast}).
The smaller scattering surface area is largely offset by the greater incident flux, rendering $C_{max}$ insensitive to $D_c$.

\begin{figure*}
 \centering
 \includegraphics[width=\textwidth]{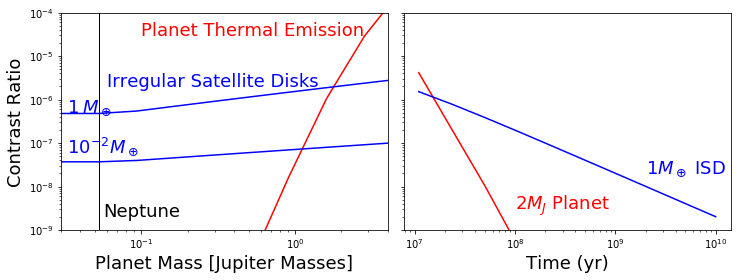}
 \caption{
Left panel plots contrast ratio vs planet mass for the same ISD and system parameters as Figs.\:\ref{mvssep} and \ref{CRvssep}. Planet is placed at 60 AU separation. 
Planet thermal emission is interpolated from 
\protect\cite{Baraffe08} (see text). 
ISD scattered light has only a weak dependence on its central planet's mass, allowing for the detection of Neptune-like planets for which thermal emission is undetectable. Right panel plots the same ISD's contrast ratio decay as a function of time for a 2 Jupiter-mass planet. ISD contrast ratios decay slower over time, allowing for a larger window of detectability.   
\label{scaling}}
\end{figure*}

\subsection{Comparison with Planetary Thermal Emission} \label{thermalcomp}
The strongest differences searching for ISDs vs planets in thermal emission is the much weaker dependence on planetary mass and time (Eq.\:\ref{CRfinal}).
As mentioned in Sec.\:\ref{secTcol}, the weak ISD contrast ratio dependence on planetary mass is due to competing effects: more massive planets have larger Hill spheres to explore for collisions (and can host smaller, higher surface-area grains), but also have higher circumplanetary velocities that speed up the rate at which phase space is explored and collisions occur.
This weak scaling is in stark contrast to current detections of super-Jovian planets in thermal emission, which fall off precipitously toward lower masses.

Second, contrast ratios decay as $t^{-1}$ (Eq.\:\ref{CRfinal}), such that the collisional timescale between the largest bodies in the cascade remains fixed to the age of the system (Sec.\:\ref{secTcol}). This scaling is also much shallower than for detections in planetary thermal emission, given that the contrast ratios are extremely strong functions of the cooling planets' temperatures.

We compare these scalings for illustration in Fig.\:\ref{scaling}, interpolating models for the planetary emission from \cite{Baraffe08}. For simplicity we assume a constant heavy element enrichment fraction $Z=0.02$.
We take a 10 Myr old, $0.5 M_\odot$ M1 host with a planet at 60 AU. 
The left panel varies the planet mass, reflecting the weak ISD dependence on this parameter in Eq.\:\ref{Cmax}, and the much steeper fall-off in planetary thermal emission.
In the right panel we plot the contrast ratio as a function of time for a 2 Jupiter-mass planet, showing the significantly shallower time decay for ISDs.

\subsection{ISDs are Typically Unresolved}

The angular size of an ISD $\theta_{ISD}$ is given by the ratio of the planet's Hill sphere radius to the target's distance from Earth,
\begin{equation}
\theta_{ISD} \approx \frac{R_H}{d} = \theta_p\Bigg(\frac{M_p}{3M_\star}\Bigg)^{1/3},
\end{equation}
where $\theta_p$ is the angular separation of the planet from the star. 
Thus, given that detections are typically made near the inner working angle at a $\theta_p$ of only a few resolution elements, the mass ratio factor $< 0.2$ implies that ISDs will typically be unresolved.
However, with large fields of view like the 10" of the IRDIS instrument on SPHERE, large-separation detections of giant planets could yield resolved ISDs.
While this may complicate the image processing, it provides a straightforward way to identify large-separation ISDs.

\subsection{ISDs are Optically Thin}

Taking the total surface area of debris $\sigma = M \xi$ as a fraction of the total ISD cross-sectional area $\pi R_H^2$ yields an optical depth $\tau$. Assuming the nominal swarm parameters in Eq.\:\ref{AMratio} then yields,
\begin{equation}
\tau \approx 0.025 \Bigg( \frac{a}{100 \text{AU}} \Bigg)^{-2} \Bigg( \frac{M_p}{M_J} \Bigg)^{-2/3} \Bigg( \frac{M_\star}{M_\odot} \Bigg)^{2/3} \Bigg(\frac{M}{M_\oplus}\Bigg).
\end{equation}

Therefore, since $M$ is expected to be $\lesssim 1 M_\oplus$ (Eq.\:\ref{Msplit} and Sec.\:\ref{masses}), ISDs should typically be optically thin, even at early times.

\subsection{ISD Scattered Light Should be Polarized}
In this optically thin limit, where single scattering dominates, one expects a strong polarization signature \citep[e.g.,][]{Kruegel03}.
Given that background stars or galaxies would not yield strong polarization, this would allow instruments with polarimetry modes, like those on GPI and SPHERE, to rule out false positives with a single epoch of observation.
Additionally, it would make it possible to more aggressively suppress unpolarized speckle noise, and enhance sensitivity.

\section{Projections} \label{projections}

With an analytic understanding of the most important scalings, we now make projections sampling parameters from reasonable ranges and optimizing the observation strategy.

\subsection{Masses} \label{masses}
While the scalings on most swarm parameters are weak (Eq.\:\ref{f_Cmax}) and expected values reasonably bracketed by those for our own populations of irregular satellites, the largest uncertainty is in the initial mass of the ISD $M_0$.
Here again we can look to our own solar system.

In steady state, the collisional cascades we have assumed above yield steep power law size distributions comparable to that observed in Jupiter's Trojan population \citep{Dohnanyi69}.
As the ISD grinds down, the eventual loss of disruptors capable of breaking up the largest bodies in the cascade causes a shallower size distribution for the biggest irregulars, as observed today \citep{Bottke10, Kennedy11}.

Debris from collisions with Saturn's largest irregular satellite Phoebe has been detected in the mid-infrared \citep{Verbiscer09, Hamilton15}, as well as in scattered optical light \citep{Tamayo14, Tamayo16Phoebe}.
However, because the shallow size distribution of the largest Saturnian irregulars suggests they have already been stranded from the collisional cascade, these observations cannot be used directly to infer $M_0$.
\cite{Kennedy11limits} searched archival data from the Spitzer infrared observatory for a more vertically extended ISD associated with the collisional cascade, but were only able to derive upper limits due to the scattered light from Saturn.
Nevertheless, the power laws of the irregular satellite size distributions around Saturn and Jupiter seem to steepen significantly for the much more numerous small $\sim 1$ km irregular satellites, consistent with the expectations of a collisional cascade.
This picture should sharpen significantly when the Large Synoptic Survey Telescope (LSST) comes online and finds many more small irregulars.
In particular, around Jupiter, LSST will go deeper in a single exposure (24.5 mag in r band) than the faintest known Jovian irregular satellites (24 mag, \citealt{Sheppard03}).

In any case, if we assume that the largest irregular satellites today, which dominate the mass in the size distribution, were initially part of the collisional cascade, we can obtain lower limits to $M_0$.
This yields minimum values from $\approx 0.5-5 \times 10^{-6} M_\oplus$ around each of the giant planets.
Given that collisional cascade masses grind down as $t^{-1}$ (see below Eq.\:\ref{M(t)}), these masses would have been boosted at early times by the ratio of the time at which the largest irregular satellites became stranded, to the time at which they were initially captured (see K11).
These quantities are highly uncertain, but if we take the stranding times around each of the Solar System giant planets estimated by K11, and the dispersal of the gas disk at 3 Myr as the time of irregular satellite capture (either through gas drag or three-body encounters), we obtain initial ISD masses in the range of $M_0 \sim 10^{-4} M_\oplus$ (Jupiter) - $10^{-2} M_\oplus$ (Neptune).
These numbers are at best rough guides, especially given that one might also expect large variations across planetary systems (with extrasolar systems estimated to typically have $\sim 1-5$ times more solid material (\citealt{Chiang13}). 
Nevertheless they provide a nominal reference scale, which we take as $10^{-3} M_\oplus$, and explore a wide range around it.

\subsection{GPIES Limits} \label{gpies}

We begin by asking whether such structures would already have been detected by direct imaging surveys if they were prevalent in young planetary systems.
As above, we draw planets with semimajor axes drawn log-uniformly from 10-200 AU \citep{Zhang18}, and masses drawn from a power-law $dN/d \log{M} \propto M^{-0.86}$ between $M_N$ and $13 M_J$ \citep{Clanton16}.
For the host stars we take the public stellar sample of GPIES, the Gemini Planet Imager Exoplanet Survey \citep{Nielsen19}, and for stellar parameters not available through VizieR catalogues \citep{vizier}, we adopt rough stellar parameters by interpolating their masses, luminosities and temperatures from values in \citep{Popper80}.
We adopt an optimistic log-uniform age distribution for the sample, from 10-100 Myr, which would tend to inflate contrast ratios. 

For the swarm parameters, we first fix the phase function $g=0.32$, the value for a Lambert sphere at maximum elongation. 
We then adopt the ranges in Table \ref{paramtable}, centered on typical values for the irregular satellites in our solar system.
The albedo of dust grains in the Phoebe ring is $\approx 0.1$ \citep{Tamayo11}
For our ISD simulation we sample from the range [0.05, 0.25], spanning roughly a factor of 2 in either direction.
Irregular satellites in the solar system vary from a few to $\sim 100$ km in size. 
We therefore choose to sample the maximum moon size $D_C$ log-uniformly from a wide range of [$10, 1000$] km. 
We sample densities uniformly from approximately the density of ice to that of silicates, and sample the transition size between the gravity and strength-dominated regimes log-uniformly in a range spanning a factor of ten around the nominal value of $100$ m adopted by K11.
Irregular satellites' characteristic orbital radius $\eta$ is well constrained between $\approx 0.2-0.4$ of their planet's Hill radius.
Finally, we sample the uncertain collisional breakup scaling factor log-uniformly from [1,10].
We summarize these adopted swarm parameter ranges in Table \ref{paramtable}.

\begin{table}
\centering
\begin{tabular}{|l|l|l|}
 Parameter      &  Range              & Description  \\ \hline
 $D_t$          &  LU[10,1000] m          & Transition between strength \& gravity regimes. \\
 $D_c$          & LU[10, 1000] km          & Largest body in the collisional cascade. \\
 $g$            & 0.32                & Phase function \\
 $Q$            & LU[0.05,0.25]     & Geometric albedo \\
 $\rho$         & U[1, 3] g/cm$^3$        & Bulk density \\
 $\eta$         & U[0.2, 0.4]            & Irregulars' orbital radius, relative to $R_H$ \\
 $f_Q$          & LU[1, 10]                & $Q_D$ scaling factor (Eq.\:\ref{QD}) \\
\end{tabular}
\caption{List of swarm parameters, descriptions, and adopted ranges used in the Monte Carlo simulations of Secs. \ref{gpies} and \ref{betapicprojections}. U and LU correspond to sampling values uniformly and log-uniformly, respectively.}
\label{paramtable}
\end{table}

The largest uncertainty is in $M_0$.
To demonstrate the GPIES should not have detected such structures, we maximize the modeled signals by adopting an optimistic initial mass in irregular satellites of 1 $M_\oplus$.
This is a natural upper limit if we assume that following formation, the giant planets accrete all leftover planetesimals within an annulus of half-width the planet's Hill radius \citep[e.g.][]{Schlichting14},
\begin{equation}
M_\text{total} \sim 1.3 \Bigg(\frac{M_p}{M_\text{Nep}}\Bigg)^{1/3}\Bigg(\frac{a}{10 \text AU}\Bigg)^{1/2} M_\oplus,
\end{equation}
where we have assumed a planetesimal disk of tens of Earth masses, as required to drive the giant planet migration thought to have occurred in the early Solar System \citep{Tsiganis05}, and that the surface density scales as $a^{-3/2}$ like the minimum-mass solar nebula.
Much of this mass would be scattered away, but it sets a natural optimistic mass scale.

\begin{figure}
 \resizebox{1.00\columnwidth}{!}{\includegraphics{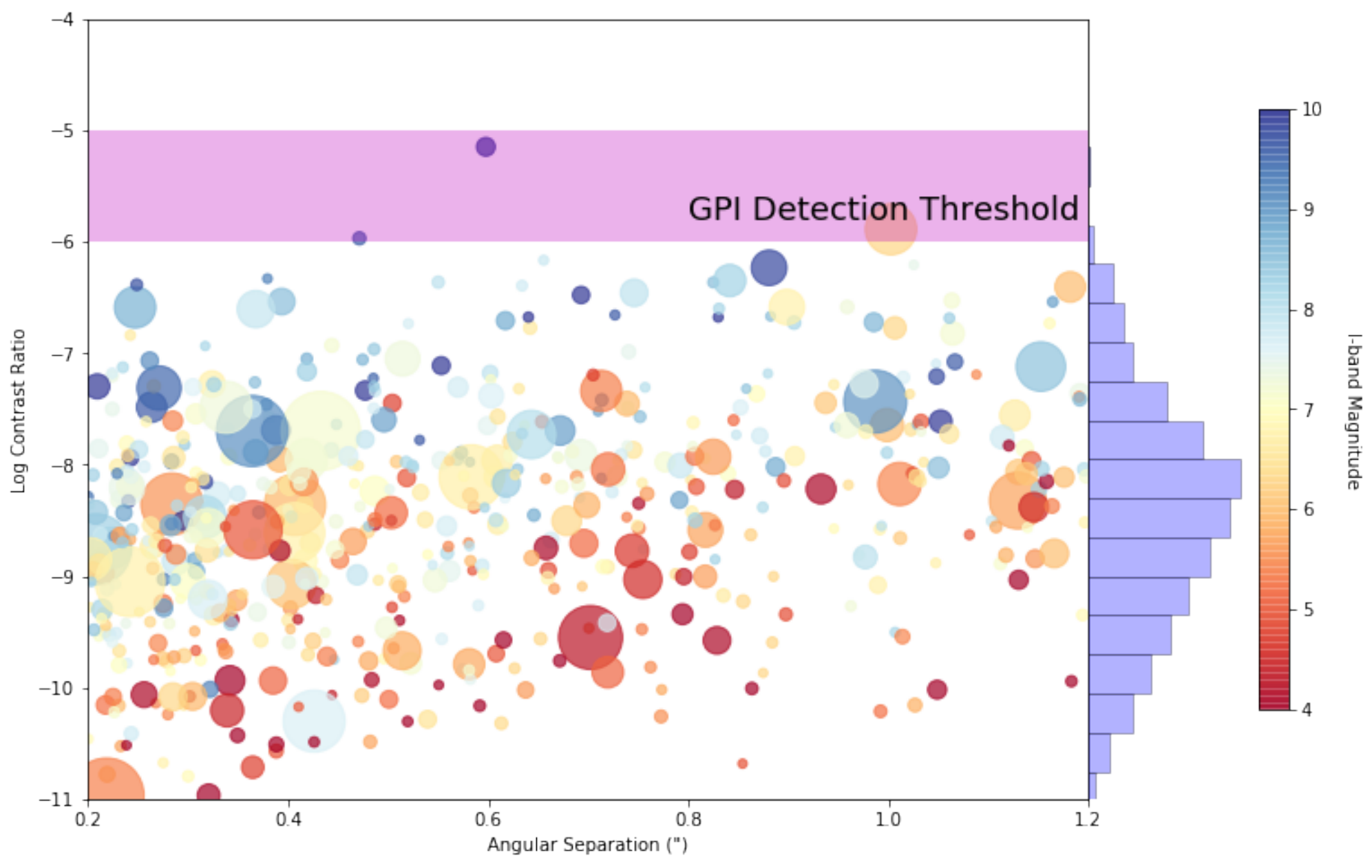}}
 \caption{Contrast ratios of ISD realizations for planets around stars in the GPIES survey sample. Even assuming optimistic ISD swarm parameters, one would not expect the GPIES survey to have detected any such structures.
 \label{figgpies}}
\end{figure}

We plot the result in Fig.\:\ref{figgpies}. 
The histogram on the right tallies detections, and the color corresponds to the star's I-band magnitude, which is the relevant wavelength band for the instrument's adaptive optics system.
It shows that GPIES not detecting ISDs was expected, even assuming optimistic ISD parameters.

There is a trend of increasing contrast ratio toward fainter stars, which can be simply understood.
Given a fixed field of view of $\approx 0.2-1.2"$, larger distances to the target star $d$ correspond linearly to larger semimajor axes. 
Thus, the contrast ratio in the collision-limited regime scales as $M_\star^{-3.71}d^{2.13}$ (Eq.\:\ref{CRfinal}), or approximately inversely with the star's bolometric flux $\sim M_\star^4d^{-2}$.

\subsection{Maximizing Contrast Ratios}

We now consider how target samples could be optimized for the detection of ISDs.
Equation \ref{CRfinal} suggests a strong preference toward lower-mass stars, with contrast ratios peaking around $\approx$ M0-M1 hosts\footnote{At masses $\lesssim 0.5 M_\odot$, the minimum particle size in the ISD becomes smaller than observation wavelengths in the optical and near-infrared, as well as the peak of the stellar emission, limiting contrast gains (Sec.\:\ref{secAMratio}).}.

As in the case of imaging planets in thermal emission, it is valuable to search for ISDs around young stars, before the swarms have ground down substantially. 
This would motivate a similar observation strategy to current direct imaging surveys of searching stars in young moving groups.
However, for ISDs, one would additionally aim for the field of view to match reasonable values of $a_{opt}$, where  the contrast ratio peaks.
The small exponent in Eq.\:\ref{aopt} constrains this range to $\sim 10-100$ AU despite large uncertainties in the initial ISD mass.

This highlights the value of a wide field of view like that of the IRDIS instrument on SPHERE (5.5"). Such an instrument can observe the closest young moving groups like the $\approx 25$ Myr old $\beta$ Pictoris moving group at $\approx 20-50$ pc from the Sun\citep{Zuckerman01, Malo14}, and still span the whole range of likely $a_{opt}$ from $\sim$ 10-100 AU.
Instruments with smaller fields of view $\approx 1"$ would be more sensitive on stars in slightly more distant young moving groups like the $\approx 8$ Myr old TW Hydra association (TWA) at $\approx 40-60$ pc from the Sun \citep{delaReza06, Ducourant14}.

\subsection{The Photon-Noise limit}. \label{photonnoise}
\begin{figure*}
 \centering
 \includegraphics[width=\textwidth]{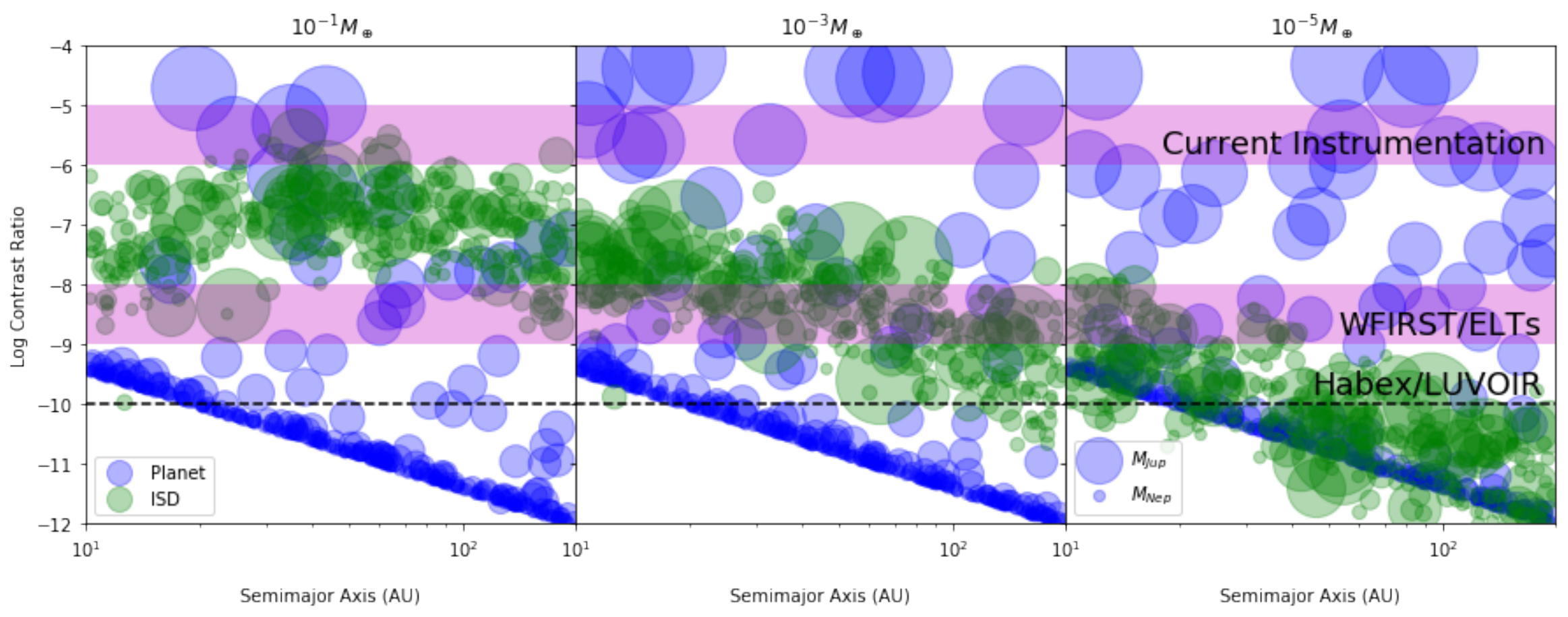}
 \caption{Contrast ratios for 300 ISD realizations around M1 stars in the $\beta$ Pic moving group. This is much larger than the number of such targets available, and just meant to statistically sample the distribution and show the probability of detection given various thresholds (magenta bands and dashed black line). Blue and green circles are the planet's and ISD's signal, respectively. Circle size is proportional to planetary mass (see legend). Each panel corresponds to a different initial mass in irregular satellites. \label{betapic}}
\end{figure*}

While current ground-based imaging is most often dominated by speckle noise, as the next generation instruments like WFIRST push toward lower contrasts, they will increasingly become photon-noise limited.

In this limit, the relevant metric of detectability is the number of photons arriving at the observer from the planet $\propto C\times L_\star$.
In the collision-limited regime, $C \propto M_\star^{-3.71}$ (Eq.\:\ref{CRfinal}).
Thus, again taking $L_\star/L_\odot \sim (M_\star/M_\odot)^4$, ISD detectability is very weakly dependent on stellar mass in this regime. 
Thus, in the photon-noise limited regime, there is no particular preference toward detecting collisionally limited ISDS around low-mass stellar targets.

Finally, for narrow fields of view in the photon-noise limited regime (as might be applicable for WFIRST), there is no strong dependence on the distance from the observer to the target star $d$. 
Here the detectability is $\propto C \times L_\star / d^2$.
Putting the same star twice as close to the observer would increase the flux by a factor of four, but since the semimajor axes probed in the fixed field of view are now twice as small, the contrast ratios $C \propto a^{2.13}$ (Eq.\:\ref{CRfinal}) have also dropped by approximately a factor of four, rendering the number of photons scattered from an ISD into the detector approximately constant.
This relative insensitivity to target distance or mass implies that follow-up observations with WFIRST-CGI of young directly imaged planets would have the added benefit of plausibly making serendipitous discoveries of lower-contrast ISDs in the same systems.

\subsection{Projections} \label{betapicprojections}

Finally, we ask what contrast ratios one would expect with a target sample optimized for finding ISDs.
For simplicity, we consider nominal M1 target stars (with mass $M_\star = 0.5 M_\odot$, luminosity $L_\star = 0.035 L_\odot$, and temperature of 3600K) in the $\beta$ Pictoris moving group, with distances drawn from a gaussian distribution centered at 35 pc and standard deviation of 10 pc, to approximately match the observed distances from $20-50$ pc \citep{Zuckerman01}.
All ISDs are assigned the moving group's estimated age of 25 Myr \citep{Zuckerman01, Malo14}.
We draw planet and swarm parameters as described in Sec.\:\ref{gpies}.
Planet thermal emission and radii (for calculating the scattered light) are derived by interpolating the same models from \citep{Baraffe08} as a function of mass, with a heavy element mass fraction $Z=0.02$ (which should  yield optimistically large planets and associated scattered light signals).
For the {\it planetary} scattered light we assume a phase function of 0.32, and a geometric albedo of 0.4.

The largest uncertainty in our model is the total initial swarm mass. 
We therefore separate out this dependency and make projections for three different initial mass scales in irregular satellites, roughly centered on the initial masses estimated for our solar system in Sec.\:\ref{masses} of $10^{-3} M_\oplus$.
We sample 300 planets with ISDs in each panel of Fig.\:\ref{betapic}.
This is a comparable sample size to the number of M dwarfs known in the $\beta$ Pictoris moving group \citep{Shkolnik17}.
Finally, we consider observing in the H-band at $\approx 1.65 \mu$m, with the wide field of view of the IRDIS instrument on SPHERE, which is able to capture the full relevant range of semimajor axes from 10-200 AU at these target distances .

We see from Fig.\:\ref{betapic} that, as expected from detections to date, several of the largest planets (large blue circles, see legend) are detectable through their thermal emission.
However, the much more common low-mass Neptunes have negligible thermal emission, and their scattered light signal in the 10-200 AU range is very weak, falling off quadratically with distance from their host star (concentrated blue line).

By contrast, ISDs (green circles) around planets of all sizes have comparable contrast ratios in each panel, as can be understood from the weak planet-mass scaling in Eq.\:\ref{Cmax}.
Planets around these optimized targets should be readily detectable by WFIRST and next-generation ELT instruments (objects in and above the corresponding magenta band) over most of the expected range in parameters.
Traversing panels from left to right, one can also see the peak at $a_{opt}$  shifting toward smaller semimajor axes. 
At $10^{-1} M_\oplus$ the peak is at $\sim 50 AU$ (Eq.\:\ref{aopt}), and by $10^{-5} M_\oplus$ has dropped to $\sim 5 AU$, so that all ISDs in the right panel are to the right of the contrast peak in the age-limited regime, where contrasts fall off quadratically like for the planets.
In all cases, the ISD scattered light signal is significantly larger than that of the planet.

One might wonder whether non-detections from coronographic observations with HST could already rule out ISDs with initial masses of 0.1 $M_\oplus$ (left panel of Fig.\:\ref{betapic}).
As noted above, HST can reach lower contrasts than the top magenta range plotted in Fig.\:\ref{betapic} around more massive, bright stars \citep{Debes19}.
However, the green ISD contrasts around these higher mass stars would be lowered $propto M_\star^{-3.71}$ (Eq.\:\ref{CRfinal}, Sec.\:\ref{photonnoise}) relative to this sample of M1 stars.
Existing observations are thus currently unable to provide strong constraints on typical ISDs around young planets.

\subsection{Direct Imaging of Exoplanet Atmospheres}  \label{atmospheres}

One might worry that bright ISDs might interfere with efforts to characterize the atmospheres of close-in planets with next-generation instrumentation that would target known planets discovered through radial velocities.
However, those planets are typically Gyrs old and, crucially, much further in.
For example, the set of radial velocity planets considered by \cite{Traub14} for WFIRST have semimajor axes of $\approx 1 AU$.
The combination of $\sim 100$ times older planets $\sim 10 \times$ farther in would result in contrast ratios a factor of $\sim 10^4$ times smaller than at the left edge of each of the panels in Fig.\:\ref{betapic} (specifically targeting M1 stars).
Thus, even with optimistic initial swarm masses, contrast ratios from ISDs would be comfortably below $10^{-10}$.
In the regime close to the host star where giant planets will be detectable through direct imaging in scattered light, ISDs would have ground down enough not to interfere with observations. 

\subsection{Competition from circumstellar debris disks}

Pushing to progressively smaller contrast ratios, one increasingly worries about interference from circumstellar debris, analogues to debris in the Kuiper belt of our solar system.
If one boosts estimated dust levels in the outer solar system by even a factor of 100, this debris would only yield contrast ratios at the level of $10^{-9}$ at tens of AU \citep{Traub14}.
However, the Herschel DEBRIS survey estimated that $28\pm 3$\% of FGK stars had circumstellar disks with contrast ratios greater than $5\times 10^{-6}$ at similar distances, with the fraction decreasing toward later stellar types \citep{Sibthorpe18}.

At lower contrast ratios, and pushing toward the lower stellar masses considered above, things become much more uncertain.
However, one would expect that the confounding effects of circumstellar debris would at least be weaker at the tens of AUs where ISDs are brightest than at the AU scales where planets would be imaged directly in reflected light.

\section{Conclusion}  \label{conclusion}
Direct imaging surveys have revealed that young super-Jupiters detectable through their thermal emission are rare \citep{Nielsen19}.
Next generation instruments will therefore target weaker scattered light signals from known, close in planets that intercept a larger fraction of starlight.
However, by increasing planets' scattering area by orders of magnitude, irregular satellite disks (ISDs) can even render far out Neptunes detectable.

We investigated ISD properties and detectability following models for their collisional evolution from \cite{Kennedy11}.
Such structures would typically be optically thin, polarized, and unresolved.
Detecting polarized point sources can rule out background sources or other false positives to high confidence in a single epoch.
One could also more aggressively suppress the unpolarized speckle noise to improve sensitivity.
We provide analytic expressions for their expected contrast ratios as a function of stellar, planetary and swarm parameters (Sec.\:\ref{secscalings}), as well as open-source code for ease of calculation and reproduction of the figures in this paper \url{https://github.com/LoicNassif/CE-Irregular-Satellites}.
This makes it straightforward to make predictions for different instruments and observation strategies.

Planets are brightest in scattered light closer to their host star where they intercept more starlight. At these distances, inside $\sim 1$ AU, ISDs will have typically ground away (Sec.\:\ref{atmospheres}), and will therefore not interfere with atmospheric characterization efforts.
By contrast, scattered light signals from ISDs are brighter at larger semimajor axes, typically peaking in the $\sim 10-100$ AU range where the exoplanet sample is sparsest.
Indeed, over a wide range of parameters, we expect that the majority of planetary signals with contrast ratios $\sim 10^{-9}-10^{-6}$ should be due to ISDs at these long periods (Sec.\:\ref{betapicprojections}).
Furthermore, the weak scalings with planetary mass and system age render a wide diversity of planets detectable (Sec.\:\ref{thermalcomp}).

Below contrast ratios of $\sim 10^{-6}$, once integration times begin to be measured in days, efficient observation strategies shift from discovery of new planets to characterization of planets discovered by alternate methods.
In this regime, wide fields of view like that of the IRDIS instrument on SPHERE would be a considerable asset for serendipitous detections. 
This would allow programs to characterize the atmospheres of known planets at $\sim 1$ AU to simultaneously detect ice giant ISDs out to $\sim 100$ AU.
This provides a complementary and compelling science case, inaccessible to other detection methods, which would elucidate the early lives of planetary systems (Sec.\:\ref{irregulars}).

\section{Acknowledgements}
We are very grateful to Kevin Schlaufman for his help and insights with stellar samples. We are also indebted to Jason Wang for numerous patient explanations, as well as Max Millar-Blanchaer, Bruce Macintosh and Eric Nielsen for useful discussions.
This research has made use of the VizieR catalogue access tool, CDS,
 Strasbourg, France (DOI : 10.26093/cds/vizier). The original description 
 of the VizieR service was published in \cite{vizier}.
 Support for this work was provided by NASA through the NASA Hubble Fellowship grant HST-HF2-51423.001-A awarded  by  the  Space  Telescope  Science  Institute,  which  is  operated  by  the  Association  of  Universities  for  Research  in  Astronomy,  Inc.,  for  NASA,  under  contract  NAS5-26555.
 This research was made possible by the open-source projects 
\texttt{Jupyter} \citep{jupyter}, \texttt{iPython} \citep{ipython}, 
and \texttt{matplotlib} \citep{matplotlib, matplotlib2}.

\bibliography{Bib}
\end{document}